\documentclass[%
 reprint,
 amsmath,amssymb,
 aps,
]{revtex4-2}

\usepackage{enumitem}
\usepackage{graphicx}
\usepackage{dcolumn}
\usepackage{bm}
\usepackage{cleveref}
\usepackage{caption}
\usepackage{subcaption}

\begin{document}


\title{Characterizing contaminant noise in barcoded perturbation experiments}

\author{Forrest Sheldon}
 \email{fs@lims.ac.uk}
\affiliation{%
London Institute for Mathematical Sciences, 21 Albemarle St. London W1S 4BS}%

\collaboration{LIMS/bit.bio Collaboration}

\date{February 10, 2023}

\begin{abstract}
Bursting cells lead to ambient RNA that contaminates sequencing data.
This process is especially problematic in perturbation experiments where transcription factors are implanted into cells to determine their effects.
The presence of contaminants makes it difficult to determine whether a factor is truly expressed in the cell.
This paper studies the properties of contaminant noise from an analytical perspective, showing that the cell bursting process constrains the form of the noise distribution across factors.
These constraints can be leveraged to improve decontamination by removing counts that are more likely the result of noise than expression.
In two biological replicates of a perturbation experiment, run across two sequencing protocols, decontaminated counts agree with bulk genomic measurements of the transduction rate and are automatically corrected for differences in sequencing.
\end{abstract}

\maketitle


\section{\label{sec:intro}Introduction}

\begin{figure*}
\begin{center}
    \includegraphics[width=\textwidth]{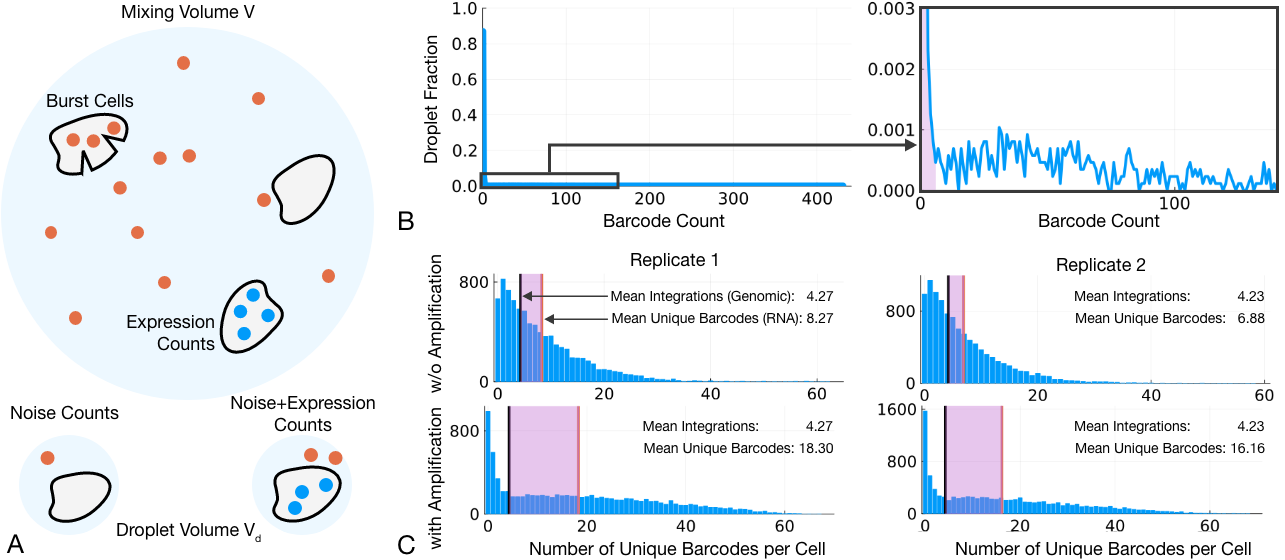}
    \caption{\label{fig:panelexpmnt}\textbf{Contaminant RNA is a significant effect in transduction experiments.} In A we show a summary of the contamination process in which bursting cells distribute free RNA throughout a large volume. For a particular gene, sequenced droplets could show just noise counts, or a combination of counts due to expression and noise. B shows example count data from replicate 1 without amplification. In the upper axes we see that most counts occupy a peak very close to zero. Zooming into small values on the y-axis, we see a second population of higher counts. The shaded region indicates that most cells contain only very few counts which are likely due to contamination. C shows that across two biological replicates and two sequencing protocols (with amplification and without amplification), the average number of unique barcodes per cell recorded from RNA sequencing exceeds the average number of integrations recorded from bulk genomic sequencing (ddPCR), confirming the presence of contaminant counts.}
\end{center}
\end{figure*}

The problem of ambient RNA in single-cell sequencing data has seen attention from the bioinformatics community~\cite{Young20,Yang20} and pipelines and tissue atlases use tools to decontaminate their counts~\cite{heumos2023best,tabsap22}.
Despite this, little work has been done to understand the properties of this contamination noise from a scientific perspective.
Primarily, this is because counts in single-cell experiments are low and observing features of the even lower contaminant distribution is difficult.

Transcription factor perturbation experiments, where cells are implanted with transcription factor genes~\cite{takahashi2006induction,moreau2016large,eguchi2016reprogramming,joung2023transcription}, conspire to make ambient RNA a more significant problem.
First, cells are exposed to greater stresses, leading them to burst more often.
Second, the controlled expression of implanted factors is often orders of magnitude higher than usual, so that the burst cells spill far more RNA into the media.
Third, the implanting process is inhomogeneous, so only a fraction of the cells will express a particular factor.
As a result, the ambient RNA is present in greater quantities and shows greater fluctuations than in typical experiments.
Finally, analysis of implanted factors, and especially combinations of factors, often relies on only a few cells as the implantation process is difficult to control and experimenters must balance the presence of several different factors.

These same features make screens an attractive arena in which to study contamination noise. 
The higher contamination levels allow us to study the distribution of contaminant counts in greater detail.
A better understanding of this distribution should translate to an improved ability to decontaminate counts.
Perturbation experiments are used to understand the underlying structure of gene interactions~\cite{meinshausen2016methods} and to identify candidate factors for reprogramming~\cite{ronquist2017algorithm}.
A precise understanding of the sources of noise in perturbation experiments is essential to making these processes reliable.

Previous work in this area also makes assumptions that are difficult to justify in perturbation experiments.
Because counts are often very low, contamination is detected as a global shift of expression values by a constant contamination vector.
Both SoupX~\cite{Young20} and  DecontX~\cite{Yang20}
 infer a constant background contamination profile that is a mixture of the different cell populations present in sequencing.
While both of these tools perform well in their domains, the underlying assumptions that each population of cells has a fixed expression profile does not apply to transcription factor screens.
In a screen, each cell may express a different subset of the implanted factors which re-configure its expression profile.
Conceivably, we could identify populations with different sets of implanted factors, but doing this requires distinguishing whether counts are the result of expression or contamination, which is precisely the task we aim to accomplish.

To address this, we examine the properties of contaminant noise from a mechanistic perspective.
We base our study on the idea that contamination noise is the result of two effects as shown in \cref{fig:panelexpmnt}A.
First, cells must burst spilling their contents into the media.
Second, the processes surrounding sorting and sequencing cause this RNA to mix over the media which is then sampled in a small volume, such as a droplet.
The contamination distribution is a balance between the fluctuations introduced by cell bursting and the dampening influence of mixing.

The experiments we examine consist of two biological replicates in which cells were transduced with a library of 82 plasmids.
Each plasmid was tagged with a short unique barcode to distinguish it from the endogenous gene~~\cite{lalli2022measuring,gosselin2016unbiased}.
Replicates underwent 10X single-cell labeling and targeted amplicon sequencing of barcodes.
In parallel, the replicates underwent sequencing with the standard gene expression library from 10X.
Gene expression data was sorted for cells using the standard cellranger pipeline.
Example count data for droplets containing cells is shown in \cref{fig:panelexpmnt}B.
The two most prominent features are the large number of counts near zero, seen in the left panel, and the population of counts extending up to very high values, seen in the zoomed axes on the right.

Due to the stochastic nature of transcription and sequencing~\cite{elowitz2002stochastic,wilkinson2020persistent,chubb2006transcriptional,herbach2019stochastic,fu2016estimating}, counts for an expressed gene should be described by a distribution, which we call the expression distribution, $P_E(e)$, that gives the probability of observing a count $E=e$ in a cell that is expressing a given gene.
We expect that the population of high counts is due to expression.

But the peak near zero also contains many non-zero counts which appear to be from a qualitatively different process.
For the gene shown in \cref{fig:panelexpmnt}A, approximately 8\% of all cells lie in this peak, with 5\% lying in the high count population.
(The remaining 87\% are zero.)
As a result, labeling all cells that contain a single count as due to expression could introduce nearly an equivalent number of false positives as true positives.

\Cref{fig:panelexpmnt}C shows distributions over cells for the number of unique barcodes seen in their counts.
Each of the replicates was also sequenced using two different protocols.
In the 10X pipeline, the cDNA is usually subjected to an initial PCR amplification step~\cite{10X}.
Because of the high counts per barcode each replicate was sequenced both with and without this amplification step, shown in the vertically aligned histograms for each  replicate.
The number of factors implanted in each cell was also measured in two ways.
The mean number of integrations was measured for each replicate using bulk genomic sequencing.
This is shown by the black vertical line in each histogram.
Additionally, the number of barcodes with at least one count was calculated for every cell.
This is shown in the histogram and the mean is shown by the red line.

The mean number of unique barcodes far exceeds the mean number of integrations, shown by the shaded region in each plot.
But the number of integrations should be an upper bound on the number of unique barcodes in a cell, due to possible duplicate integrations and silencing of integrated genes.
Without the initial amplification step, the number of unique barcodes is roughly double the average number of integrations.
This is exactly what we would expect if the counts in the peak near zero in \cref{fig:panelexpmnt}A were due to contamination.

The amplification step has a strong effect on contamination.
In the lower histograms, the unique number of barcodes increases to nearly 4 times the number of integrations.
While making analysis more challenging, the higher contamination rate can help us observe features of the contaminant distribution.

We would like a method that not only removes contaminant counts, but also that gives us a measure of trust in this process to be put to use in downstream analysis.
This should also correct for differences in sequencing, bringing the two protocols of the replicates into alignment.
Our approach will be to derive a connection between the contamination distribution $P_C(c)$ and expression distribution $P_E(e)$ and then fit a mixture model to the counts of the form,
\begin{equation}\label{eqn:mix}
    P_S(s) = fP_{E+C}(s) + (1-f)P_C(s)
\end{equation}
where $f$ is the probability that a factor is implanted and expressed in a cell.
As shown in \cref{fig:panelexpmnt}A the sequencing distribution $P_S(s)$ indicates that the observed count $s$ is the sum of expression counts and contaminant counts with probability $f$, or consists only of contaminant counts with probability $1-f$.
Regions in which $fP_{E+C}(s) > (1-f)P_C(s)$, roughly corresponding to the unshaded region in \cref{fig:panelexpmnt}B, are where counts are more likely the result of expression than noise.
The relative magnitudes of these probabilities also tell us how confident we should be in this labeling.
Intuitively, a count of 100 is more likely the result of expression than noise, while a count of 5 may be ambiguous.

In the following section we derive a general form for the contaminant distribution as a function of the expression distribution of the gene and three parameters related to bursting, mixing and sequencing respectively.
In the results section we examine evidence that these parameters are shared amongst all genes and leverage this to improve fitting of lowly expressed genes.
Applying this to a transcription factor screen allows us to infer which factors are expressed by a cell and remove those that are unlikely.
Without the need for transforming or normalising counts, this process brings the two sequencing protocols into close alignment and corrects for the increased contamination due to amplification.

\section{The Contaminant Distribution}

\begin{figure}
    \begin{center}
  \includegraphics[width=\columnwidth]{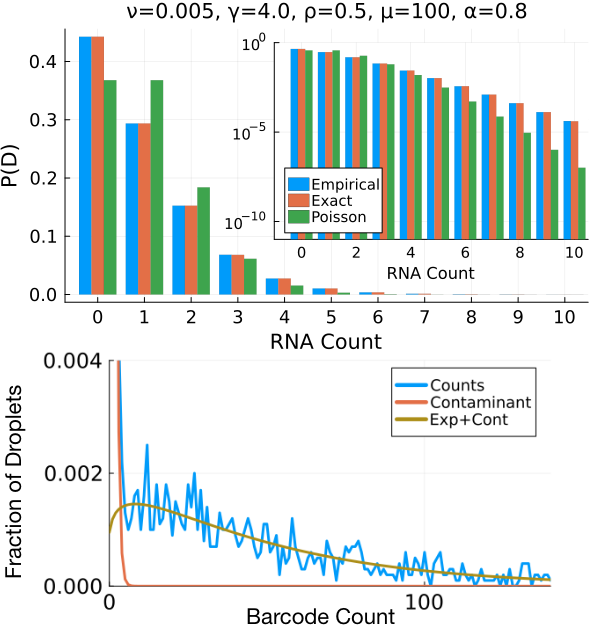}
  \caption{\label{fig:Exacthist} \textbf{A comparison between exact, empirical and Poisson contaminant distributions.} The exact distribution from \cref{eqn:gC} is plotted against an empirical distribution of $10^7$ samples and the Poisson distribution with the same mean $\gamma\mu\nu\rho$. The Poisson distribution underestimates both the zero counts and the tail. When the mixture distribution in \cref{eqn:gmix} at $f=0.1$ is fit to simulated data, the contaminant and expression distributions are inferred.
  }
  \end{center}
\end{figure}

We first derive a relationship between any expression distribution and the resulting contaminant count distribution.
The results are compatible with a wide variety of possible expression distributions, including parametric models like the negative binomial~\cite{hilbe2014modeling}, empirical distributions~\cite{feuerverger1984statistical} and non-parametric Bayesian estimates~\cite{bintayyash2021}.
In order to keep notation simple, we will at first assume that every cell expresses the gene of interest (\emph{i.e.} $f=1$ in \cref{eqn:mix}).
After establishing the form of the contaminant distribution, a simple parameter change will allow us to generalize to an inhomogeneous population.

The derivation involves several summations over a random number of terms.
These sums can be concisely expressed in terms of generating functions~\cite{feller1968introduction}.
Given a positive integer random variable $X\sim P(X)$ its probability generating function is
\begin{equation}\label{eqn:genfunc}
    g_{X}(z) = \sum_{x=0}^\infty P(X=x)z^x.
\end{equation}
For many distributions, $g_X$ is a familiar function.
For the Bernoulli random variable with parameter $q$, the generating function is a line $b_q(z) = 1 + q(z-1)$.
For the Poisson random variable with parameter $\gamma$, the generating function is an exponential $p_\gamma(z) = \exp(\gamma(z-1))$.
We will use the notations $p_\gamma$ and $b_q$ for the Poisson and Bernoulli generating functions in what follows.
The generating function is in one-to-one correspondence with the distribution with probabilities and moments given by expanding $g_X(z)$.

Generating functions simplify working with sums of random variables.
For two random variables $A$ and $B$, the generating function of their sum is, $g_{A+B}(z) = g_A(z)g_B(z)$.
As a consequence, the sum of a random numbers of terms is given by a composition of generating functions.
For example, if we sum $N$ copies of a random variable $X$, where $N$ is also a random variable, the generating function of the sum is $g_N(g_X(z))$.

The sorting and sequencing process involves separating cells into droplets.
Few droplets contain a cell~\cite{10X}.
As such, the droplets that contain cells may have been drawn at different times and be minimally correlated with each other.
We thus neglect dynamical effects and regard each droplet as being drawn independently.

We model contaminant counts as resulting from a sequence of processes shown in \cref{fig:panelexpmnt}.
\begin{itemize}[noitemsep]
    \item Each cell expresses a number of transcripts $E_i$.
    \item $B$ of these cells burst.
    \item The resulting $R=\sum_{i=1}^B E_i$ molecules of RNA mix over a volume $V$.
    \item A droplet of volume $V_d$ is drawn from the volume, containing $D$ of these molecules.
    \item $C$ of the $D$ molecules are picked up in sequencing.
\end{itemize}
We would like to obtain the distribution of $C$ (or equivalently it's generating function).

Each of these steps can be modeled with a familiar distribution, leading to a succession of compositions of generating functions.
Leaving the distribution of $E$ arbitrary, but with generating function $g_E(z)$, 
\begin{itemize}[noitemsep]
    \item If cell bursting is rare and independent, we can assume that the number of burst cells $B$ is Poisson distributed with parameter $\gamma$.
    \item As a result, the number of molecules of RNA $R$ is a random sum and has generating function $g_R(z) = p_\gamma(g_{E'}(z))$, where $g_{E'}(z)$ is the generation function of the expression distribution.
    \item We assume that each of these molecules is independently distributed over the volume $V$ and is thus picked up in the droplet with probability $\nu = \frac{V_d}{V}$. The number of RNA molecules in the droplet has the generating function $g_D(z) = g_R(b_\nu(z))$.
    \item Finally in the sequencing step, each of the $D$ molecules in the droplet is captured with probability $\rho$, giving the contaminant generating function $g_C(z) = g_D(b_\rho(z))$.
\end{itemize}

Assembling these pieces and simplifying, the resulting generating function is
\begin{equation}\label{eqn:gC}
    g_C(z) = p_\gamma\left(g_{E'}\left(b_{\nu \rho}(z)\right)\right)
\end{equation}
where $p_\gamma$ is the Poisson generating function and $b_{\nu\rho}$ is the Bernoulli generating function shown above.

\Cref{eqn:gC} expresses the relationship between the number of captured contamination counts and the expression distribution of the cell in terms of three parameters: the cell bursting rate $\gamma$, the volume ratio $\nu$,  and the sequencing capture rate $\rho$. 
The benefit of introducing these extra parameters is that they do not depend on any particular gene.
As a result, they should take nearly equivalent values across all genes, which can be leveraged in decontamination.

In \cref{fig:Exacthist} we show a comparison between simulation data, the exact distribution described by \cref{eqn:gC} and a Poisson distribution with the same mean.
The contaminant count distribution initially decreases more rapidly than in the Poisson distribution but with a more slowly decaying tail.
This is a result of fluctuations introduced by the cell bursting rate $\gamma$.
It is also simple to show that if we take the cell bursting rate $\gamma\to\infty$ with the product $\nu\gamma\rho$ held constant, $C$ approaches a Poisson distribution with the parameter $\nu\gamma\rho\mu$ where $\mu$ is the mean of the expression distribution.
This holds for any expression distribution and is derived in the supplemental material.

Lastly, in an experiment, only a fraction of cells $f$ are successfully transfected and express a gene.
The substitution $\gamma\to f\gamma$ includes this inhomogeneity in the contaminant distribution.
The mixture distribution in $\cref{eqn:mix}$ can be then be expressed in terms of these generating functions as
\begin{equation}\label{eqn:gmix}
    g_S(z) = f g_E(z)g_C(z) + (1-f)g_C(z)
\end{equation}
where $g_C$ depends on $f$ through $g_C(z) = p_{f\gamma}\left(g_E\left(b_{\nu \rho}(z)\right)\right)$. 

For the most commonly used expression distribution, the negative binomial~\cite{hilbe2014modeling}, the generating function is
\begin{equation}
    n_{\mu,\alpha}(z) = \left(1-\alpha \mu (z-1)\right)^{-1/\alpha}.
\end{equation}
Using this for the distribution of the number of molecules in an expressed gene, the expression distribution after sequencing is
\begin{equation}\label{eq:gE}
    g_E(z) =n_{\mu, \alpha}(b_\rho(z))= n_{\mu\rho,\alpha}(z)
\end{equation}
and the contaminant distribution is
\begin{align}\label{eqn:gCNB}
    g_C(z) =& p_{f\gamma}(n_{\nu\mu\rho, \alpha}(z))\nonumber\\
    {} =& \exp\left(\gamma\left(\left(1-\alpha \nu\mu\rho (z-1)\right)^{-1/\alpha} - 1\right)\right).
\end{align}

The distribution given by \cref{eqn:gmix} is now a function of five parameters, $\nu$ and $\gamma$ which relate to contamination, and $\mu\rho$, $\alpha$, and $f$ which relate to expression.
We have combined $\rho$ and $\mu$ into a single parameter as they only occur as a product in $g_S$.
This is also why we introduced the notation $E'$ in \cref{eqn:gC}, which refers to the distribution of the expression distribution before sequencing.
In \cref{eq:gE} we have replaced $E'\to E$, to indicate the inclusion of the sequencing step and the replacement $\mu\to \rho\mu$.

The benefit of this approach is that $\nu$ and $\gamma$ should take the same value across all genes as the bursting and mixing processes are not gene dependent.
The distribution suggests that in the low contamination limit, contamination should be approximately Poisson distributed with the product $\nu\gamma$ equal across all genes.
For higher contamination, each parameter should converge to fixed values separately.
In the next section we will examine evidence for this in a perturbation experiment.

In order to do this, we need to fit these distributions to count data.
We found that the most reliable method consisted of of a combined method of moments to obtain initial conditions~\cite{wasserman2004all} followed by maximum likelihood.
The probabilities were obtained from the generating function in \cref{eqn:mix} by forward auto-differentiation~\cite{tucker2011validated} using Julia's TaylorSeries.jl package and optimization was carried out using Julia's Optim.jl implementation of L-BFGS.
This process is summarized in \cref{fig:panelinf}A and described more fully in the supplemental material.
An implementation, with a graphical user interface, data and notebooks constructing figures in this manuscript, is available on github at \url{https://github.com/forrestsheldon/BinTF}.

An example fit to simulated data is shown in \cref{fig:Exacthist} for 10,000 counts.
Estimates for the parameters appear unbiased and the variance of the relative error converges as $1/N$.
For a number of observations on the order of a perturbation experiment, we should be able to reliably infer the component distributions, so long as expression $\rho\mu$ is sufficiently large.

\begin{figure*}
    \begin{center}
  \includegraphics[width=\textwidth]{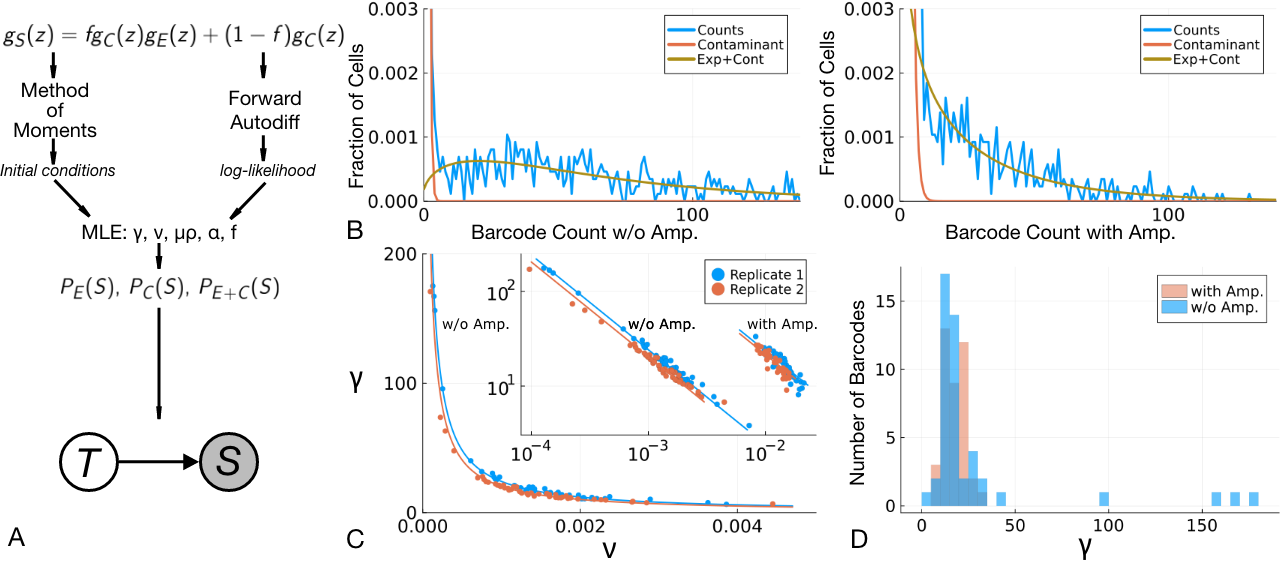}
  \caption{\label{fig:panelinf} \textbf{A summary of the inference procedure and parameters.} A gives the steps in fitting the contaminant mixture distribution. The distributions obtained from the fit can become components of graphical models that connect sequencing data to phenotype changes. B shows fits obtained for the same barcode with and without an amplification step. The increased number of low counts can be seen in the plot with amplification. C shows that the fit parameters $\gamma$ and $\nu$ collapse onto the curve $\nu\gamma = const.$ with the same constant for all barcodes. The inset shows both sequencing protocols on logarithmic axes where the curve becomes a line. The higher contamination increases the constant, but the value of $\gamma$ appears similar across sequencing protocols and replicates. In D, the histogram of $\gamma$ shows that the values are tightly concentrated across both sequencing protocols. }
  \end{center}
\end{figure*}

\section{Results}

Transcription factors were first filtered for sufficient counts.
From $\approx 8000$ cells, we selected factors that appeared in at least 200 cells with a count of at least 10.
This included 49 factors in each replicate without amplification and 41 and 42 factors in replicates 1 and 2 respectively, with amplification.


Example fits and the distribution of the fit parameters $\gamma$ and $\nu$ are shown in \cref{fig:panelinf}.
The components of the mixture distribution form a close approximation to the count distributions for nearly all genes selected (a single gene in each replicate was removed due to a poor fit).

Plotting $\gamma$ and $\nu$ across these genes in \cref{fig:panelinf}C, we see that they collapse onto a curve of the form $\nu\gamma = c$ in both replicates when sequenced without amplification.
This is just as we would expect from the Poisson limit, where $\nu\gamma$ is the only meaningful contaminant parameter and the individual values of $\gamma$ and $\nu$  are unconstrained.

In the inset of \cref{fig:panelinf}C, we show both sequencing protocols on log-scales.
The shift to the right for the amplified protocol reflects a larger value of $\nu\gamma$ and an increase in contamination. 
The product $\nu\gamma$ connects the mean contamination with the barcode specific parameters $f$ and $\rho\mu$.
The increase is driven by an increase in $\nu$ with the amplification step.

The histogram in \cref{fig:panelinf}D shows that the values of $\gamma$ across both replicates and sequencing protocols are tightly concentrated
Larger values occur the lower contamination protocol, as we would expect from the Poisson limit.
But, while $\gamma$ converges to the same value in both protocols, $\nu$ is sequencing dependent, and takes distinct values in each protocol.

While the distribution introduces $\nu$ as a geometrical parameter (the ratio between the capture and mixing volumes), it's role is similar to that of $\rho$, in that it reflects how contaminant molecules are collected.
An increase in the droplet volume is equivalent to an increase in the probability that molecules are captured and so $\nu$ acts as a contaminant specific $\rho$.
As a result, we conclude that the amplification step is increasing the collection of contaminants more rapidly than counts due to expression.
We will address this more carefully in the discussion.
Though $\nu$ is sequencing dependent, the fits suggest that for a single replicate and sequencing protocol, both $\gamma$ and $\nu$ are at least tightly concentrated and appear to take nearly the same value across all of the sequenced barcodes.

\section{Application}

\begin{figure*}
    \begin{center}
  \includegraphics[width=\textwidth]{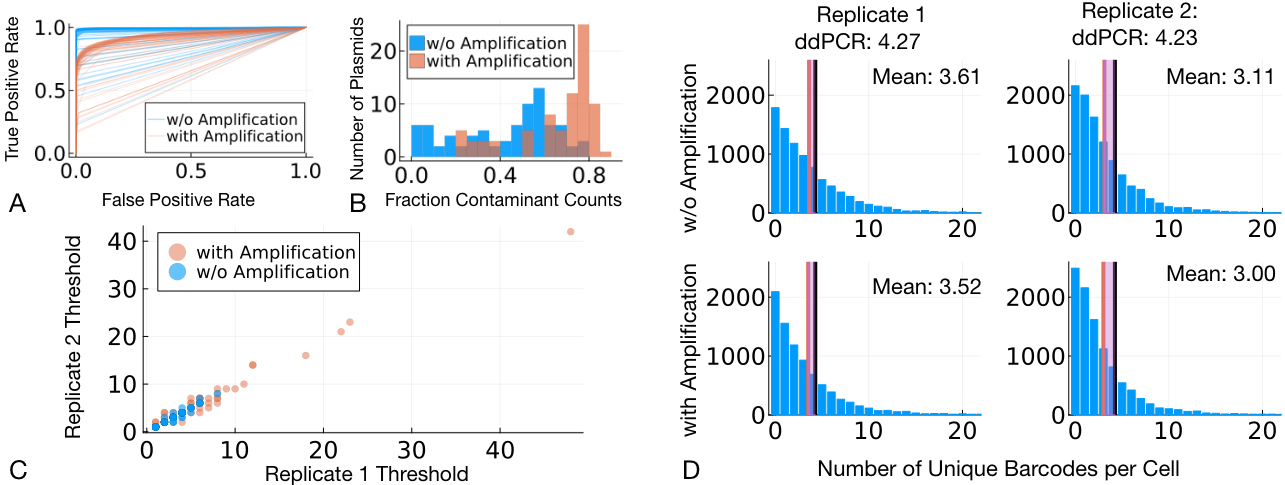}
  \caption{\label{fig:panelapp} \textbf{Decontamination and error rates.} A shows the ROC curves for all barcodes in replicate 1. The increased contamination with amplification leads to worse classification (and thus decontamination) performance. B shows the fraction of nonzero counts that are more more probably due to contamination, according to the fit distributions. For many barcodes over half of all counts are likely due to contamination. C compares the inferred thresholds between replicates. The inferred thresholds span a wide range. Finally D shows the corrected copy number distributions after below threshold counts are removed. The average numbers of implanted factors are in close agreement between sequencing protocols and below ddPCR as expected. }
  \end{center}
\end{figure*}

We can apply the fit distributions to decontaminating counts.
By using the values of $\nu$ and $\gamma$ from the high-count factors to constrain their values in the low count factors, we can obtain contamination estimates for otherwise difficult to fit count distributions.

Within each replicate and protocol we form the estimates $\hat{\gamma} = \text{median}(\gamma)$ and $\hat{\nu} = \text{median}(\nu\gamma)/\hat{\gamma}$.
Their values are then fixed in the fit of all factors.
Additionally, $\alpha$ was found to be highly concentrated across factors and was similarly fixed to it's median.
With these set, it remains to fit the gene specific parameters $\rho\mu$, and $f$.

This can be accomplished through either maximum likelihood or a method of moments, both of which are detailed in the supplemental material and implemented in the accompanying github repository.
This process is robust to changes in the number of counts and converges for all factors in the experiment, though other options are also discussed below.

The resulting distributions can be used to infer whether a cell is likely to contain a particular transcription factor.
Framing this as a simple graphical model as shown in \cref{fig:panelinf}A, we want to infer whether a transcription factor was successfully implanted or not ($T\in\{0, 1\}$) given a count value $S$.
Identifying $P(T=1) = f$, we can express this in terms of our model parameters as,
\begin{equation}\label{eqn:gm1}
    P(T=1\mid S) = \frac{fP_{E+C}(S)}{fP_{E+C}(S) + (1-f)P_C(S)}.
\end{equation}
Keeping only counts for which $P(T=1\mid S) > P(T=0\mid S)$ allows us to calculate a factor specific threshold $\theta$, beyond which counts should be kept as likely due to expression,
\begin{equation}
    \theta = \min_s \{s \mid fP_{E+C}(S=s) \ge (1-f)P_C(S=s)\}.
\end{equation}
This is just the crossing point of the two components of the fit distribution.
Discarding counts below this threshold allows us to decontaminate our dataset.

The fit distributions clearly show role of increased contamination in the protocol with an initial amplification step before the decontamination process.
The ROC curves in \cref{fig:panelapp}A show that at any fixed false positive rate, the runs without amplification achieve a higher true positive rate.
Similarly, \cref{fig:panelapp}B shows that among nonzero counts, the inferred fraction due purely to contamination can be the majority of all counts.
This is true independent of sequencing protocol, but is exacerbated by amplification.

As a result, inferred thresholds, shown in \cref{fig:panelapp}C, are far higher in the replicates with amplification.
When these are used to remove contaminant counts, the updated distributions for the number of unique factors come into close agreement across both replicates and protocols, and with genomic readings, shown in \cref{fig:panelapp}D.
Within the same replicate, the number of unique factors differ by approximately 3\% across the two protocols.
The inferred thresholds are thus able to correct for the increased contamination with only small losses in the number of implanted factors.
The mean number of unique factors is also below the genomic values from ddPCR, as we would expect due to the possibility of multiple copies and silencing of implanted factors.

\section{Discussion}

The distributions in \cref{fig:panelexpmnt}, when compared with genomic sequencing, show that contamination from ambient RNA is a serious problem in barcode sequencing experiments.
An immediate criticism may be that this is only a feature of this experiment, and that a more careful treatment of the cells could reduce the presence of contamination.
While we cannot survey all experiments, we note that similar disagreement between bulk genomic and single-cell RNA readings have been observed in all perturbation experiments that we have examined.
The most likely causes appear to be high expression of exogenous factors, increased bursting of stressed cells and smaller library size.
As a result, analysis is often presented with far more false positive than true positive counts.
This stands in contrast to single-cell RNA sequencing experiments where false negatives are thought to be a more serious problem~\cite{svensson2020droplet,jiang2022statistics,qiu2020embracing,sarkar2021separating}.

\Cref{fig:panelinf}C shows that at least one parameter of the noise distribution is shared amongst all factors.
The evidence for two shared factors is more uncertain.
The spread of parameters in the replicates without amplification is consistent with the Poisson limit of the distribution.
With amplification, the contaminant parameters appear to concentrate, but the extent to which this will continue is not clear.
It is certainly plausible that the a particular factor could cause a cell more stress than another and make it more susceptible to bursting, leading to a larger $\gamma$.
On the other hand, fluctuations in the counts will always lead to some spread in the measured values.
In either case, the values seem sufficiently tightly concentrated to assume that a fixed value among all genes is a close approximation to the true distribution.
The results of the decontamination procedure appear to confirm this.

While $\gamma$ takes on nearly the same value across replicates and sequencing protocols, $\nu$ is altered as an initial cDNA amplification increases it's value.
The model suggests $\nu$ as a purely geometrical quantity but it essentially represents the probability of capture for contaminants.
It is thus natural that sequencing protocols such as amplification would impact it's value.

What is surprising is that this does not affect the capture probability $\rho$ in the same way.
By comparing barcodes in the same replicate across the two sequencing protocols, we can see that $\rho$ decreases with amplification.
This could be anticipated, as an increase in the number of molecules from amplification could result in the decrease in the chance of any individual molecule's capture.
But for rare molecules like contaminants, amplification appears to selectively increase their chance of being captured in sequencing.
This is precisely what we would hope for in using amplification, but as we see here it may have the effect of selectively increasing noise in a sample.


Finding a robust method to fit these mixture is a difficult problem.
Maximum likelihood tends to fail  for mixture models as the likelihood function can become unbounded and display multiple maxima~\cite{robertson1970bias}.
While the method presented in \cref{fig:panelinf}, relying on numerical automatic differentiation, leads to reasonable results it is admittedly messy and we can forsee a few directions for research based on this.

First, the mixture in \cref{eqn:gmix} is expressed in terms of generating functions and so calculation of the maximum likelihood objective must be handled numerically.
Generating functions can be fit to the empirical generating function~\cite{feuerverger1984statistical} but for mixture distributions where the value $f$ may be be small, this process lead to poor results as one component dominated the fit.
Methods analogous to expectation maximization should be possible but we are not aware of any specific methods that have been developed.

\Cref{eqn:gmix} has two additional complications that may be of interest.
First, the mixture parameter $f$ also appears within the parameters of the contaminant distribution and so the probabilities are nonlinear in $f$.
Second, parameters are shared between both components, even for a single factor, and so the fit of each component cannot be performed independently.
Fitting methods that work directly with the generating function, and which are adapted to mixtures and sums of random variables like these would likely produce faster and more robust results once developed.

We also chose to perform an initial fit on each factor independently to obtain an estimate of the contaminant parameters, and then used this estimate to fix their value in subsequent fits for decontamination.
Fitting all genes with a shared contaminant parameters simultaneously would be another option, but this is computationally expensive.
Are there methods akin to coordinate descent that handle a mixture of shared and independent parameters efficiently?
The use of shared measurement models~\cite{sarkar2021separating} in sequencing would require similar methods and will be a fruitful direction of research.

Finally, there are several immediate applications of this work that we have not yet explored.
While here we have relied on access to gene expression data for cell calling, this is not necessarily the case in all experiments.
The barcode distributions can be immediately applied to cell calling by modifying the mixture distribution to include the possibility of empty droplets.
An expressed factor in a droplet implies the presence of a cell and an extension of the graphical model in \cref{fig:panelinf}A can be applied to cell calling.

A more complex application is in ranking combinations of transcription factors that drive acquisition of a particular phenotype.
The mixture of false positive counts and rare acquisition of phenotypes makes extracting information from these experiments difficult and having measures of the reliability of results is essential.
The component distributions we fit here can become the basis for a fully probabilistic treatment of perturbation experiments which we hope to explore in forthcoming work.

The data and code required to perform this analysis, along with a graphical user interface for labeling cells with their incorporated factors is available on github at \url{https://github.com/forrestsheldon/BinTF}.

\acknowledgements{FS recognizes the support of the LIMS/bitbio collaboration as well as the aid of several collaborators at bitbio. Among these, Ania Wilczynska, Maria Dermit and Andrew Knight were especially essential in bringing this work to its completion by making data available and for offering many helpful conversations. }

\bibliography{cellprog}

%
\pagebreak
\onecolumngrid
\appendix

\section{Poisson Limit of $g_C(z)$}

We consider the $\gamma\to\infty$ limit of $g_C(z)$ such that the product $\nu\gamma\rho = A$ is held constant. Under this assumption,
\begin{equation}
    g_C(z) = \exp\left(\gamma \left(g_E(1 + \frac{A}{\gamma}(z-1))-1\right)\right).
\end{equation}
As $\gamma\to \infty$ we have
\begin{align}
    g_E(1 + \frac{A}{\gamma}(z-1)) &\approx g_E(1) + g_E^\prime(1) \frac{A}{\gamma}(z-1)\\
    {} &\approx 1 + \mu_E\frac{A}{\gamma}(z-1)
\end{align}
which follows from $g_E(1) = 1$ and $g_E^\prime(1) = \mu_E$, the mean expression count. Substituting this gives the final result,
\begin{equation}
    g_C(z) = \exp(A \mu_C(z-1))
\end{equation}
showing that in the large $\gamma$ limit, the noise will approach a Poisson distribution with parameter $\nu\gamma\rho\mu_E$ for any expression distribution.

\section{Threshold and Error Derivations}

To derive these we introduce a two variable joint distribution for variables $S, \, T$ which govern the sequenced count and expression of an implanted transcription factor, respectively. The variable $S$ can take values in the natural numbers $\{0, 1, 2\dots\}$ corresponding to observed counts while the variable $T$ governs whether those counts are due to noise ($T=1$), or due to the combination of noise and and cell expressing the factor $T=1$.

We can encode these assumptions by factoring the distribution as,
\begin{align}
    P(S, T) &= P(T)P(S\mid T)\\
    P(T = 1) &= f, \quad P(T=0) = 1-f\\
    P(S=s\mid T&=1) = P_{C+E}(s)\\
    P(S=s\mid T&=0) = P_C(s).
\end{align}
With these definitions, this reproduces the distribution used in the main text as,
\begin{align}
    P(S=s) &= \sum_{t\in\{0,1\}}P(S=s, T=t)\\
    {}&= fP_{C+E}(s) + (1-f)P_{C}(s).
\end{align}

We can now formulate the labeling as a problem of inference. Calculating the probability of an expression state given an observed count,
\begin{equation}
    P(T\mid S) = \frac{P(T,S)}{P(S)} = P(S\mid T)\frac{P(T)}{P(S)}
\end{equation}
we have that the probability that $T=1$ is greater than $T=0$ when,
\begin{align}
P(T=1\mid S=s) &> P(T=0\mid S=s)\\
P_{C+E}(s)f &> (1-f)P_C(s).
\end{align}
If the distribution $C$ decays more quickly than $C+E$, this will be true for all $s$ greater than some minimum $\theta$ which we define as the threshold.


\section{Experimental Summary}

Human iPSC cells were transfected with a library of 82 plasmids. Each plasmid encodes the protein sequence of a single transcription factor (TF) and is tagged with a short unique barcode. The TFs integrate into the genome, and the expression of their mRNA is induced by the addition of an antibiotic. Cells having undergone cellular reprogramming are dissociated and subjected to single cell labeling using a 10x Chromium instrument. The resulting GEM (gel-beads in emulsion) were then subjected to targeted amplicon sequencing of barcodes using Illumina MiSeq v3. In parallel, the same GEMs were subjected to standard gene expression library sequencing using the 10x GEX Single Cell 3’ Reagent Kit and sequenced on a NovaSeq 6000. The GEX data was analysed using the standard cellranger pipeline and empty droplet and cell-containing droplet calls were used to separate out cells signal from noise. The barcode counts per cell were assessed based on a perfect match of the sequenced barcodes to the barcode library. The resulting data was in the form of two matrices of barcode counts per droplet for empty and cell-containing droplets.

\section{Preprocessing}
Barcode counts were first filtered for transcription factors with sufficient counts to have the fit converge. In this case we required that barcodes have at least 200 counts with a value greater than 10. Fits were convergent for many factors with lower counts, but for initial development and understanding properties of the noise distribution we restricted out attention to the most high quality data. Of 82 transcription factors in the experiment, 49 barcodes were kept.

\section{Fitting}
Fitting the mixture distribution to counts consists of two steps:
\begin{enumerate}
\item An approximate method of moments calculation is used to set the initial values for $\nu, \gamma, \mu, \alpha$ and $f$ for each gene by using an imposed threshold $t$.
\item The probabilities for the mixture distribution are calculated using an automatic differentiation package (TaylorSeries.jl) and then fit using the L-BFGS method with bounding box. An implementation can be provided upon request.
\end{enumerate}
These steps are detailed below.

\subsection{Method of Moments Calculation}

In order to calculate the initial conditions for maximum likelihood, we pick a threshold $t=10$ and assume that all counts $c\ge t$ are due to expression and all counts $c < t$ are due to noise.
While these samples are actually from the conditional distribution given the threshold, they give reasonable estimates for an initial condition, which is all we require.

Partitioning the counts into expression counts $\{x_i\}_{i=1}^{N_{ex}}$ and noise counts $\{y_i\}_{i=1}^{N_{no}}$ we have the moment conditions for the expression distribution,
\begin{align}
\langle E\rangle &= \mu\\
\langle E^2\rangle &= \mu + (1+\alpha)\mu^2
\end{align}
which give
\begin{align}
\mu &= \langle E \rangle \\
\alpha &= \frac{\langle E^2\rangle - \langle E \rangle^2 - \langle E \rangle }{ \langle E \rangle^2}.
\end{align}
The mixture parameter is calculated as
\begin{equation}
f = \frac{N_{ex}}{N_{no} + N_{ex}}.
\end{equation}
These can then be used for the noise distribution moment conditions which are,
\begin{align}
\langle C\rangle &= f\gamma\mu\nu\\
\langle C^2\rangle &= f\gamma\mu\nu + (f\gamma\mu\nu)^2 + f\gamma (\mu\nu)^2(1+\alpha).
\end{align}
These can be inverted to obtain
\begin{align}
\nu &= \frac{\langle E\rangle}{\langle C \rangle}\frac{\langle C^2\rangle - \langle C\rangle^2 - \langle C\rangle}{\langle E^2\rangle - \langle E\rangle}\\
\gamma &= \frac{\langle C\rangle}{f\mu\nu} = \frac{1}{f} \frac{\langle C\rangle^2}{\langle E\rangle^2} \frac{\langle E^2\rangle - \langle E\rangle}{\langle C^2\rangle - \langle C\rangle^2 - \langle C\rangle}\\
\end{align}
Once calculated these become the initial conditions for the maximum likelihood calculation.

\subsection{Maximum Likelihood}

The counts $\{s_i\}_{i=1}^N$ are used to calculate the maximum likelihood objective,
\begin{equation}
L(\nu, \gamma, \mu, r, f) = \sum_{i=1}^N \log P_{seq}(s_i)
\end{equation}
where the probabilities $P_{seq}(c_i)$ are calculated using the TaylorSeries.jl package to expand
\begin{equation}
g_{S}(z) = fg_{C}(z)g_{E}(z) + (1-f)g_{C}(z)
\end{equation}
This is optimized using Julia's Optim.jl package to find the parameter values for each gene.

\subsection{Constraining Global Parameters}

\begin{figure}
\centering
\begin{subfigure}{.5\textwidth}
  \centering
  \includegraphics[width=0.9\textwidth]{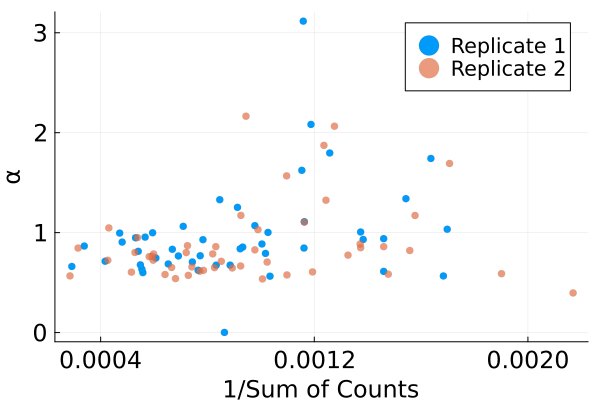}
  \caption{Without Amplification}
\end{subfigure}%
\begin{subfigure}{.5\textwidth}
  \centering
  \includegraphics[width=0.9\textwidth]{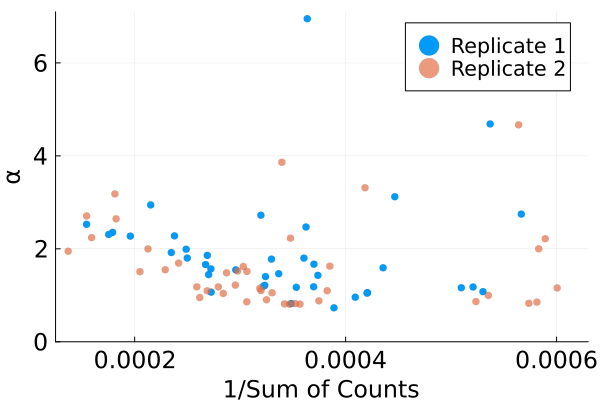}
  \caption{With Amplification}
\end{subfigure}
\caption{\label{fig:alpha} \textbf{Finite size scaling of $\alpha$.} As the number of counts collected increases, $\alpha$ shows signs of concentrating to a single value across all the fit barcodes.}
\end{figure}

The outcomes of these initial fits are used to constrain the values of $\gamma$ and $\nu$.
These are shown in the histogram in the main text and were set to the median value taken across all barcodes. 
Due to instability in the final fits, we also chose to set $\alpha$ to a single value in each replicate and sequencing protocol.
In \Cref{fig:alpha} we have plotted $\alpha$ against the inverse sum of counts in each condition.
The infinite sample size limit is at 0 on the horizontal axis.
As the number of counts increases, the barcodes show signs of concentrating along the vertical axes.

The results are certainly not as unambiguous as we might like, especially in the conditions with amplification where the trend is less obvious.
But we note that the results appear very similar across both replicates, and indicate that $\alpha$ is also sequencing dependent.
Amplification appears to increase the dispersion of the counts, as we might expect.
We have used the median to estimate the asymptotic value of $\alpha$, which appears to be a good estimate without amplification.
With Amplification this is likely an underestimate and a weighted median or regression based estimate might be a better option.
As the probabilities and threshold are mostly slowly varying with $\alpha$, we chose to stick with the median for the sake of simplicity but there are certainly other options available.

\section{Final Method of Moments}

After fixing the values of $\nu$, $\gamma$ and $\alpha$, we can also use a method of moments to fit the gene specific parameters. In this we will set $\rho=1$. The mean count, second moment and variance are:
\begin{align*}
    \langle S\rangle =& f\langle E+C\rangle + (1-f)\langle C\rangle\nonumber\\
    {} =& f\langle E\rangle + \langle C\rangle\\
    \langle S^2\rangle =& f\langle E^2 + 2EC + C^2\rangle + (1-f)\langle C^2\rangle\nonumber\\
    {}=& f\langle E^2\rangle + 2f\langle E\rangle\langle C\rangle + \langle C^2\rangle\\
    \langle S^2\rangle -\langle S\rangle^2 =& f\langle E^2\rangle - f^2\langle E^2\rangle + \langle
    C^2\rangle - \langle C\rangle^2.
\end{align*}
We have the following collection of moments of the component distributions,
\begin{align*}
    \langle E\rangle =& \mu & \langle E^2\rangle =& \mu + (1+\alpha)\mu^2\\
    \langle C\rangle =& \nu\gamma f\mu & \langle C^2\rangle =& \langle C\rangle^2 + \langle C\rangle(1+(1+\alpha)\nu\gamma)
\end{align*}
Inserting these, we have the following two moment equations,
\begin{align}
    f\mu =& \frac{\langle S\rangle}{(1 + \nu\gamma)}\\
    \mu =& \frac{\frac{\langle S^2\rangle - \langle S\rangle^2}{\langle S\rangle}(1 + \nu\gamma) + \frac{\langle S\rangle}{(1 + \nu\gamma)} - (1+\nu\gamma)}{(1+\alpha)(1+\nu^2\gamma)}
\end{align}
which give $f$ and $\mu$. The method of moments fit is also implemented in the github repository \url{https://github.com/forrestsheldon/BinTF}.

\end{document}